\let\myover=\over    
\documentclass[12pt,a4paper]{article}
\usepackage{amssymb,epsfig}  

\def\e{{\rm e}}

\newcommand{\be}{\begin{equation}}
\newcommand{\ee}{\end{equation}}

\def\l{\left(}
\def\r{\right)}

\def\e{{\rm e}}

\newcommand{\bg}{\begin{gather}}
\newcommand{\eg}{\end{gather}}

\newcommand{\bea}{\begin{eqnarray}}
\newcommand{\eea}{\end{eqnarray}}                             

\begin{document}
\let\over=\myover   
\topmargin 0pt
\oddsidemargin=-0.4truecm
\evensidemargin=-0.4truecm

\renewcommand{\thefootnote}{\fnsymbol{footnote}}


\newpage
\setcounter{page}{1}
\begin{titlepage}
\vspace*{-2.0cm}

\vspace*{0.5cm}
\begin{center}
{\Large \bf Gravity waves from inflating brane} \\
{\Large \bf or} \\
{\Large \bf Mirrors moving in adS$_5$
}
\vspace{0.5cm}


{\large D.S.~Gorbunov
\footnote{E-mail: gorby@ms2.inr.ac.ru}, V.A.~Rubakov
\footnote{E-mail: rubakov@ms2.inr.ac.ru}, S.M.~Sibiryakov
\footnote{E-mail: sibir@ms2.inr.ac.ru} 
\vspace*{0.4cm}}


{\small\em Institute for Nuclear Research of the Russian Academy of Sciences, 
Moscow 117312, Russia}
\end{center}
\begin{abstract}
We study tensor perturbations in a
 model 
with inflating Randall--Sundrum-type brane embedded in  
five-dimensional anti-de~Sitter (adS$_5$) bulk.
In this model, a natural {\it in}-vacuum of gravitons is
the vacuum defined in static adS$_5$ frame.
We show that this vacuum is, in fact, the same as the
{\it in}-vacuum defined in the frame with de Sitter
(dS$_4$) slicing, in which the brane is at 
rest. Thus, 
during inflation, gravitons on and off the brane
remain in their vacuum state.
We then study  graviton creation by the brane on which inflation
terminates at some moment of time. We mostly consider
gravitons whose wavelengths at the end of inflation
exceed both the horizon size and the adS$_5$ radius.
Creation of these  gravitons is 
dominated by (zero mode)--(zero mode) Bogoliubov
coefficients and, apart from an
overall  factor, 
the spectrum 
of produced gravitons
is the same as in four-dimensional theory.
``Kaluza--Klein'' gravitons
are also produced, 
but this effect is subdominant.
Graviton spectra at somewhat higher momenta
are also presented for completeness.
\end{abstract}

\end{titlepage}
\renewcommand{\thefootnote}{\arabic{footnote}}
\setcounter{footnote}{0}
\newpage


\section{Introduction and summary}

A special property of brane-world models with infinite extra 
dimension(s), namely, the Randall--Sundrum (RS)
model~\cite{RS}
and its generalizations (for a review see, e.g., 
Ref.~\cite{Rubakov:2001kp}) is the existence of soft bulk 
gravitons with continuum spectrum of four-dimensional masses.
This property is of interest in the cosmological context,
as in principle it may affect the spectrum of
long-wavelength perturbations created at inflation, 
and therefore the large
scale structure and CMB anisotropy. In inflationary models, 
perturbations which are of long-wavelengths today,  
are created  from zero-point fluctuations of high initial
momenta, the corresponding  ``Kaluza--Klein'' (KK)
zero-point fluctuations  are 
generically not suppressed near the brane at early times, 
and thus may not 
be negligible. Also, if soft KK gravitons were abundantly
produced at inflation and survived until recent epoch, 
they would leave footprints
in
CMB anisotropy.

The simplest inflationary brane-world model 
is that of inflating RS brane embedded in the 
five-dimensional anti-de~Sitter (adS$_5$) bulk.
Naturally, considerable effort has been put into
the study of the generation of
perturbations in this 
model~\cite{Company,Hawking:2000kj,Langlois:2000ns},
in particular, the tensor 
perturbations~\cite{Hawking:2000kj,Langlois:2000ns}.
The latter analyses utilized either  brane-based
approach~\cite{Langlois:2000ns} or adS/CFT 
correspondence~\cite{Hawking:2000kj}. 
However, the most straightforward 
approach
to the study of the generation of tensor perturbations,
which we adopt in this paper,
is to solve the five-dimensional linearized
Einstein equations with appropriate boundary conditions on
the brane, and calculate the  Bogoliubov coefficients.
Such a calculation is similar to the study
of particle creation by moving mirrors: in static
adS$_5$ coordinates, the three-brane 
moves~\cite{Kraus:1999it}
and acts as a mirror (with Neumann boundary conditions)
for five-dimensional
gravitons.
The calculation of this sort enables one, at least 
in principle, to evaluate the spectrum of created gravitons
(not only in the zero-mode final state, 
as in Ref.~\cite{Langlois:2000ns}, but also in continuum states)
and also to find the relevant correlators
(not only at the de~Sitter stage of the brane evolution,
as in Ref.~\cite{Hawking:2000kj}, but also after inflation).

In this paper 
we first consider an inflating brane with de Sitter (dS$_4$)
intrinsic geometry, which is a boundary of a part of adS$_5$.
We address the following issue. The natural 
graviton vacuum in adS$_5$,
cut by the brane, is the state which does not contain
five-dimensional gravitons moving towards the brane
(we refer to this state, somewhat loosely, as
adS$_5$-vacuum).
This definition needs 
qualification, as it involves the choice of a coordinate
frame in adS$_5$. The appropriate frame is static, so the
brane moves in this frame. From the point of view of an
observer residing on the brane, another coordinate frame
in adS$_5$ is appropriate, in which the brane is at rest. 
The latter frame, which corresponds to  dS$_4$
slicing of adS$_5$,
defines its own graviton vacuum,
which may or may not coincide with the adS$_5$-vacuum.
In Section 2 we find that, in fact, 
the vacuum defined with respect to dS$_4$ slicing 
is the same as
 adS$_5$-vacuum.
This means that 
the fact that the {\it in}-vacuum is defined in static adS$_5$
frame does not add any physics above and beyond that inherent in
dS$_4$ slicing; gravitons on and off the brane
remain in their vacuum state.
This situation is in clear contrast to the case of
a mirror accelerating in Minkowski space-time.
Our result in a sense contradicts expectations expressed 
in Ref.~\cite{Alexander:2001ic}, 
but we think it is natural.
Indeed, consider a brane on which matter in its
{\it ground state} 
has energy density in excess to the fine-tuned RS value.
Then there is no reason to suspect that the system would excite
itself, i.e., it is natural that quantum fields (including 
gravitons) remain in their ground state, and the 
expansion of the brane is precisely 
de~Sitter\footnote{In the case of
accelerated mirror in Minkowski space-time, there is external force
that accelerates the mirror. 
Such external force is absent in our case.}.

Technically, our calculation involves the evaluation of the
Bogoliubov coefficients relating two sets of graviton 
wave functions: one set consists of waves moving towards the 
brane, with frequency decomposition defined in the
static adS$_5$ frame, while another 
set~\cite{Garriga:2000bq,Langlois:2000ns,Karch:2001ct}
has fixed frequencies in the frame corresponding
to dS$_4$ slicing. We find that the
non-trivial Bogoliubov coefficients vanish, i.e., the wave functions
which are negative frequency  in one frame, do 
not have positive-frequency components in another frame.

In Section 3 we proceed to 
  calculate
the graviton creation by the brane on which inflation
terminates at some moment of time.
We are interested mostly in 
gravitons
whose wavelengths at the end of inflation are larger than 
both the 
dS$_4$ horizon size $H^{-1}$
and adS$_5$ radius $k^{-1}$. We  argue 
in Appendix 1
that 
for all realistic adS$_5$ radii, only this part of
the spectrum is relevant for the generation of 
CMB anisotropy at  interesting angular scales.  

We find no 
surprizes. The creation of large wavelength gravitons is 
dominated by (zero mode)--(zero mode) Bogoliubov
coefficients and, apart from the
overall enhancement factor \cite{Langlois:2000ns}, the spectrum 
of produced gravitons
is the same as in four-dimensional theories. The 
continuum KK gravitons 
 are also produced, but their number is 
suppressed. As anticipated in Ref.~\cite{Langlois:2000ns},
KK part of initial
zero-point fluctuations gives negligible
contributions into production of gravitons in both
zero and continuum modes. Hence, the spectrum of
large wavelength gravitons, created due to inflation on
the brane, is indistinguishable from the spectrum
predicted by four-dimensional theories.

If the Hubble parameter on the inflating brane, $H$,
is much larger than the inverse adS$_5$ radius
$k$, there is a range of graviton spatial momenta
at the end of inflation, $p_{phys}$, in which the
gravitons are superhorizon at this moment of time,
but their wavelengths are smaller than the adS$_5$
radius, $H \gg p_{phys} \gg k$. For completeness, 
we consider creation of these gravitons in
Appendix 4. We find that the spectrum of created zero-mode gravitons 
gets damped
at these momenta. One peculiar feature of this case is that
the creation of KK gravitons in the final state
dominates over the production of zero-mode gravitons,
in contrast to the case of soft gravitons studied
in the main text. Still, the energy density emitted
into the bulk in the part of the spectrum
$H \gg p_{phys} \gg k$
 is a small fraction of the energy density 
on the brane. The emission of gravitons into the
bulk is a small effect at least for gravitons which are
superhorizon by the end of inflation.

\section{dS$_4$ mirror moving in adS$_5$}

\subsection{{\it in}-vacuum in adS$_5$}

The adS$_5$ metric in static coordinates is
\be
  ds^2 = \frac{1}{(kz)^2}(dt^2 - d{\bf x}^2 - dz^2)
\label{frame1}
\ee
A spatially flat three-brane moving in these coordinates
is described by a world surface
\be
  t = t (\eta) \; , \;\;\;\;  z = z(\eta)
\nonumber
\ee
where $\eta$ is the intrinsic time coordinate on the
brane.

Throughout this paper we consider tensor perturbations,
i.e., write the perturbed metric as follows,
\be
ds^2 = \frac{1}{(kz)^2}[dt^2 - 
(\delta_{ij} + h_{ij})d x^i dx^j - dz^2]
\nonumber
\ee
where $h_{ij}$ are transverse and traceless. The perturbations
 are 
decomposed in spatial Fourier modes
\be
   h_{ij} (t, {\bf x}, z) =
\mbox{e}^{i {\bf p}{\bf x}}~ e_{ij}~ \Phi (t,z;p)
\nonumber
\ee
where $e_{ij}$ are constant transverse-traceless polarization
tensors, and $\Phi$ obeys the Klein--Gordon equation for minimal
massless scalar field in adS$_5$~\cite{Company,Langlois:2000ns},
\be
\left( - \frac{\partial^2}{\partial z^2}
+ \frac{3}{z} \frac{\partial}{\partial z}
+ \frac{\partial^2}{\partial t^2} + p^2 \right)
\Phi = 0
\label{KGstatic}
\ee
With $Z_2$ orbifold symmetry imposed across the brane,
the boundary condition for $\Phi$ is Neumann on the brane,
\be
    \partial_n \Phi = 0
\label{boundarygen}
\ee
where $\partial_n$ denotes the derivative normal to the
brane.

The {\it in}-vacuum in 
adS$_5$ is defined by choosing the basis of incoming waves 
(falling towards the brane) which have fixed frequencies
in the static frame. 
The solutions to eq.~(\ref{KGstatic}), 
which are incoming waves at large $z$
and have fixed frequencies, are\footnote{These
solutions do not
obey the Neumann boundary conditions (\ref{boundarygen})
on the  brane.
However, as long as one considers
asymptotic past,
one can use wave packets composed of
these functions, and disregard this subtlety.}
(up to normalization, which is unimportant 
for our purposes)
\be
   \phi_{in-adS,~m}^{(+)} = \mbox{e}^{i\omega t} z^2
H_2^{(1)} (m z)
\label{adSset}
\ee
and
\be
   \phi_{in-adS,~m}^{(-)} = (\phi_{in-adS,~m}^{(+)})^{*}
= \mbox{e}^{-i\omega t} z^2
H_2^{(2)} (m z)
\label{adSset2}
\ee
where  $H_2^{(1,2)}$ are the Hankel functions, the 
continuous parameter $m$ is positive, $m>0$, 
and
\be
    \omega = \sqrt{m^2 + p^2}
\nonumber
\ee
These solutions behave at large $z$ as follows
\be
     \phi_{in-adS,~m}^{(\pm)} 
=-\sqrt{\frac{2}{\pi m}} \cdot z^{3/2}
\mbox{e}^{\pm i(\omega t + m z-\pi/4) }
\label{adS-asym}
\ee
As a side remark, we note  that the normalization condition
is, generally speaking,
\be
\int~dz~\sqrt{g}~g^{00}~i(\phi_{m'}^{*} 
\partial_t \phi_m  - \partial_t \phi_{m'}^{*} 
\phi_m ) = \delta_{m', m}
\label{measure}
\ee
The factor $z^{3/2}$ in eq.~(\ref{adS-asym})
is precisely what is needed to normalize 
these solutions to delta-function
$\delta(m'-m)$ with the weight $\sqrt{g} g^{00}
\propto z^{-3}$.

\subsection{dS$_4$ slicing}

The frame in which dS$_4$ brane is static is obtained
by introducing coordinates $\eta < 0 $ and $\xi$, 
related to $t$ and $z$ as follows
\begin{eqnarray}
   t &=& \eta \cosh \xi 
\nonumber \\
   z &=& - \eta \sinh \xi
\nonumber
\end{eqnarray}
In these coordinates, the adS$_5$ metric is
\be
   ds^2 = \frac{1}{(k\sinh \xi)^2}\left[\frac{1}{\eta^2}
\left(d\eta^2 - d{\bf x}^2 \right) - d\xi^2\right]
\label{frame2}
\ee
The dS$_4$ brane is at constant $\xi$,
\be
  \xi = \xi_B
\label{xib}
\ee
i.e. in static coordinates it moves along the straight line
\begin{eqnarray}
 t &=& \eta \cosh \xi_B
\nonumber \\
   z &=& - \eta \sinh \xi_B
\label{dSmotion}
\end{eqnarray}
It is clear from eqs.~(\ref{frame2}) and (\ref{xib}) that
the induced metric on the brane is indeed de~Sitter, 
the Hubble parameter on the brane is
\be
    H = k \sinh \xi_B
\label{hubble1}
\ee
and the
coordinate
$\eta$ is the conformal time in dS$_4$.

In the new coordinates, the Klein--Gordon
equation is
\be
\left[ \left(\frac{\partial^2}{\partial \eta^2}
-\frac{2}{\eta} \frac{\partial}{\partial \eta}
\right) - \frac{1}{\eta^2}
\left(\frac{\partial^2}{\partial \xi^2}
-\frac{3 \cosh \xi}{\sinh \xi} \frac{\partial}{\partial \xi}
\right) + p^2 \right] \Phi = 0
\label{dSKG}
\ee
Since the brane is static, 
the boundary condition 
(\ref{boundarygen}) takes a simple form,
\be
   \partial_\xi \Phi\biggr|_{\xi = \xi_B} = 0
\label{boundaryxi}
\ee
One solution to these equations is
the bound state (zero mode), $\phi_{0} (\eta)$,
which depends only on $\eta$.
The zero mode which in infinite past has
negative frequency {\it with respect
to time} $\eta$ is
\be
   \phi_{0}^{(-)} = k^{\frac{3}{2}} \sinh \xi_B
\frac{C_1}{\sqrt{p}} \mbox{e}^{-ip\eta}\left(\eta -
\frac{i}{p} \right)
\label{dSzeromode}
\ee
where the normalization factor $C_1$ is determined by
eq.~(\ref{measure}) written in $(\eta, \xi)$ frame,
\be
\int~\frac{d \xi}{(k \sinh \xi)^3 \eta^2}~i(\phi_{\kappa'}^{*} 
\partial_{\eta} \phi_\kappa  - \partial_\eta \phi_{\kappa'}^{*} 
\phi_\kappa ) = \delta_{\kappa', \kappa}
\label{measure2}
\ee
Here $\kappa$ labels the solutions to eqs.~(\ref{dSKG}), 
(\ref{boundaryxi}); for the moment $\kappa = \kappa^{\prime}=0$. 
One finds from eq.~(\ref{measure2})
\be
 C_1^2 \cdot 2(\sinh \xi_B)^2 \int_{\xi_B}^{\infty}~
\frac{d \xi}{\sinh^3 \xi} = 1
\label{normalization-of-zero-mode}
\ee
The quantity $C_1$ is precisely the same as that
introduced in Ref.~\cite{Langlois:2000ns}. 
In the case of
slow inflation, $H \ll k$ (and hence $\xi_B \ll 1$), 
one has $C_1 =1$, while at $H \gg k$ one finds
$C_1 = \sqrt{3H/2k}$.

Let us now consider non-zero modes of eq.(\ref{dSKG}).
The variables decouple, so 
the modes have the form
\be
  \phi_{\kappa} (\eta, \xi)
= \psi_{\kappa}(\eta) \cdot \chi_{\kappa} (\xi)
\label{dSfull}
\ee
where $\psi_{\kappa}$ obeys 
\be
 \left(\frac{\partial^2}{\partial \eta^2}
-\frac{2}{\eta} \frac{\partial}{\partial \eta}
 + \frac{\kappa^2 + \frac{9}{4}}{\eta^2}   
+ p^2 \right) \psi_{\kappa} = 0
\label{dStime}
\ee
whereas the spatial part of the wave function obeys
the boundary condition (\ref{boundaryxi}) and equation
\be
\left(\frac{\partial^2}{\partial \xi^2}
-\frac{3 \cosh \xi}{\sinh \xi} \frac{\partial}{\partial \xi}
+ \kappa^2 + \frac{9}{4}
\right) \chi_{\kappa} = 0
\label{dSspace}
\ee

Equation (\ref{dStime}) may be reduced to the Bessel equation.
Its solution which is negative-frequency at large negative $\eta$
is 
\be
\psi^{(-)}_{\kappa}(\eta) =
 \frac{\sqrt{\pi}}{2}~ k^{\frac{3}{2}} ~ 
\mbox{e}^{\frac{\pi \kappa}{2}} ~ |\eta|^{\frac{3}{2}}
~ H_{i\kappa}^{(2)} (p\eta)
\nonumber
\ee
At large negative $\eta$ it behaves as follows
\be
 \psi^{(-)}_{\kappa} = -\frac{1}{\sqrt{2p}}~ k^{\frac{3}{2}}
~ \eta ~ \mbox{e}^{- ip\eta - i \pi/4}\;.
\nonumber
\ee
The normalization factor is chosen in such a way that
\be
\frac{1}{k^3 \eta^2}~i(\psi^{(+)}_\kappa 
\partial_\eta \psi^{(-)}_\kappa
- c.c.) = 1
\nonumber
\ee
(where $\psi^{(+)}_{\kappa} = (\psi^{(-)}_{\kappa})^{*}$).
The factor $k^{-3} \eta^{-2}$ here comes from the measure
in (\ref{measure2}).

The spatial wave functions $\chi_\kappa$ are considered 
in detail in Appendix 2. Their form is not important in this 
section. It is worth pointing out, however, that
there are no bound states except for the zero mode,
and that the parameter $\kappa$ is continuous and starts 
from zero.

\subsection{Vacuum in dS$_4$ slicing coincides with 
adS$_5$-vacuum}

Let us see that the wave functions $\phi^{(-)}_{\kappa}$,
which are negative frequency with respect to time $\eta$,
are also negative frequency with respect to time $t$.
This will mean that the graviton vacua
defined in the two frames (\ref{frame1}) and
(\ref{frame2}), respectively, are in fact the 
same\footnote{There is a subtlety here. The dS$_4$ slicing
(\ref{frame2}) and correspondingly the wave functions
$\phi_{\kappa}$ are defined only in the past
light cone, $z < -t$. As long as physics at or near 
the brane is considered, this subtlety is irrelevant.}.

We are going to calculate the Bogoliubov coefficients
between the sets $\phi^{(-)}_{\kappa}$
and \linebreak
$\phi_{in-adS, m}^{(+)} \equiv (\phi_{in-adS, m}^{(-)})^{*}$.
These are given by scalar products similar to the left hand 
sides of eqs. (\ref{measure}) or (\ref{measure2}),
but now involving the functions
$\phi^{(-)}_{\kappa}$
and 
$\phi_{in-adS, m}^{(+)}$.
We prefer to integrate over the hypersurface of fixed $\eta$,
and then take the limit $\eta \to - \infty$. Hence
the Bogoliubov coefficients of interest are proportional to
\be
v_{\kappa,m}^{dS-adS} \propto  
\int_{\eta \to - \infty}~\frac{d \xi}{(\sinh \xi)^3 \eta^2}~
(\phi_{\kappa}^{(-)} \partial_{\eta} [\phi_{in-adS,m}^{(+)}]^{*}
-  [\phi_{in-adS,m}^{(+)}]^{*} \partial_{\eta}  \phi_{\kappa}^{(-)})
\label{intdsads}
\ee
where the adS$_5$ wave functions are expressed through coordinates
$\eta$ and $\xi$,
\be
[\phi_{in-adS,m}^{(+)}]^{*} \equiv
\phi_{in-adS,m}^{(-)} = \mbox{e}^{- i \omega \eta \cosh \xi}
(\eta \sinh \xi)^2 H_2^{(2)}(-m \eta \sinh \xi)
\nonumber
\ee
Let us first note that at large negative $\eta$,
both the zero mode $\phi_0^{(-)}$ and non-zero
modes $\phi_{\kappa}^{(-)}$, corresponding to dS$_4$ slicing,
behave in a similar way,
\be
   \phi_{\kappa}^{(-)} = \eta \mbox{e}^{-ip\eta} 
\chi_{\kappa}(\xi)
\nonumber
\ee
At large negative $\eta$, the integrand in eq.~(\ref{intdsads})
rapidly oscillates,
and we may use the saddle-point approximation.
The saddle point $\xi_s$ is finite, so at large $|\eta|$ we use
the asymptotic form of the Hankel function 
$H_2^{(2)}(- m \eta \sinh \xi)$, see also (\ref{adS-asym}).
In this way we obtain
\begin{eqnarray}
v_{\kappa,m}^{dS-adS} \propto 
 \sqrt{|\eta|}~
\int_{\eta \to -\infty} & & \frac{d\xi}{\sinh^{3/2} \xi} \chi_{\kappa}(\xi)
\mbox{e}^{- i \omega \eta \cosh \xi + im \eta \sinh \xi}
\mbox{e}^{- ip\eta} \times
\nonumber \\
& & \left[- \omega \cosh \xi + m  \sinh \xi +
p + O(\eta^{-1})\right]
\label{tmp1}
\end{eqnarray}
The saddle point is at 
\be
 \tanh \xi_s = \frac{m}{\omega}
\nonumber
\ee
that is
\be
  \sinh \xi_s = \frac{m}{p} \; ,\;\;\;
  \cosh \xi_s = \frac{\omega}{p}
\nonumber
\ee
At this point, however, the pre-factor vanishes,
up to possible corrections decaying as $\eta^{-1}$.
Thus
\be
v_{\kappa,m}^{dS-adS} = 0
\nonumber
\ee
as promised.

A remark is in order.
Should we calculate $u$-coefficients, relating
components $\phi^{(\pm)}_{\kappa}$
and 
$\phi_{in-adS, m}^{(\pm)}$
of the {\it same} sign of frequencies, 
the pre-factor would 
not vanish at the saddle point, and we would obtain the result
which is finite as $\eta \to -\infty$ (the overall factor
$\sqrt{|\eta|}$ in ($\ref{tmp1}$) would cancel $|\eta|^{-1/2}$
which would come from the Gaussian integration near the saddle point).
This may be viewed as a cross check of our calculation.

\section{Gravity waves from inflation on the brane}

\subsection{A model for the background}

The main purpose of this paper is to calculate
the spectrum of gravitons created by the brane
which inflates for long period of time, and then
enters the Friedmann regime. As long as one is interested in 
gravitons which are superhorizon by the end of inflation,
it suffices to use the following simplified model
for the background. We consider 
the brane which inflates until 
some moment of time $\eta_0 <0$, and then does not evolve 
at all. In a similar model in four dimensions, the calculation
of graviton creation is very simple, and is presented in
Appendix 3. Of course, its result coincides with 
the result of realistic calculations which take into account
evolution of the Universe after inflation.

\begin{figure}[htb]
\begin{center}
\epsfig{file=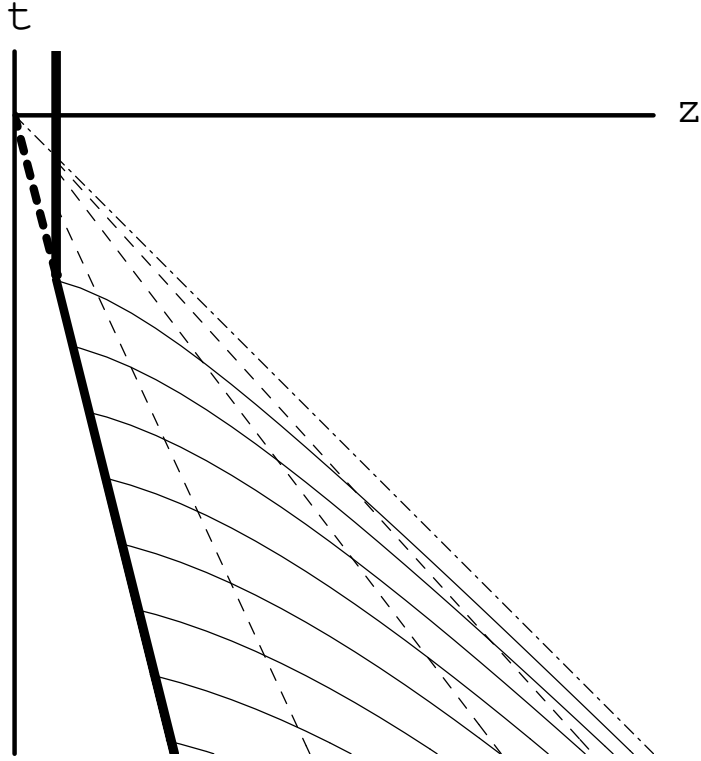}
\end{center}  
\caption{Brane motion in static coordinates: thick solid line
represents the world surface of the brane, 
dash-dotted line denotes the past light
cone; thin solid (dashed) lines correspond to surfaces of fixed
$\eta$ ($\xi$).}
\label{plot}
\end{figure}

In static coordinates the trajectory
of the brane is the straight line (\ref{dSmotion})
for $\eta < \eta_0$, and then another straight line
parallel to the $t$-axis, 
\be
z = z_B = |\eta_0| \sinh \xi_B
\nonumber
\ee
as shown in Figure~\ref{plot}. 
The brane after inflation is Minkowskian, with the scale
factor on the brane being
\be
       a(z_B) = \frac{1}{k |\eta_0| \sinh \xi_B}
\nonumber
\ee 
Of course, one could rescale $\eta$ in such
a way that the scale factor on Minkowski brane is equal to 1,
but it is convenient to keep $\eta_0$ arbitrary, 
and make sure, as a cross
check, that the results depend on physical quantities.
Namely, the physical momentum at the end of inflation
(and on Minkowski brane)  is related to the coordinate momentum
$p$ as follows,
\be
  p_{phys} = \frac{p}{a(z_B)} = p k |\eta_0| \sinh \xi_B
\nonumber
\ee
Comparing this expression with eq. (\ref{hubble1}), we find 
that the modes which are superhorizon at the end of inflation,
 obey
\be
     \frac{p_{phys}}{H} \equiv  p |\eta_0|  \ll 1
\label{peta0}
\ee
We will be interested mostly in modes whose physical momentum at
the end of inflation is small compared to the inverse 
adS$_5$ radius, see Appendix 1. 
In terms of the coordinate momentum, this means
\be
     \frac{p_{phys}}{ k} \equiv p  |\eta_0| \sinh \xi_B
 \ll 1
\label{pk}
\ee
In view of these relations, it will be convenient to
consider $|\eta_0|$ as a small parameter of the problem.

We note in passing that at $H \gg k$,
the spatial momenta of superhorizon modes
may be such that the relation (\ref{pk}) is violated.
Though the corresponding range of momenta,
$H \gg p_{phys} \gg k$, is not particularly interesting
in the cosmological context, we consider this range 
in Appendix 4 for completeness.

\subsection{ In-out Bogoliubov coefficients: generalities}

We are going to calculate the Bogoliubov coefficients
which relate the graviton wave functions corresponding to
{\it in}-  and {\it out}- vacua. The wave functions of the
{\it in}-vacuum have been already described. In our simplified
model with Minkowski brane at late times, the wave functions of
{\it out}-vacuum are the standard RS wave functions
(modulo factors accounting for the fact
that the brane is placed at $z = z_B$).
Namely,
the normalized zero mode is
\be
   \varphi_0^{(\pm)} = z_B ~\frac{k^{\frac{3}{2}}}{\sqrt{p}}~
\mbox{e}^{\pm ipt}
\label{RSzero}
\ee
Non-zero modes are
\be
\varphi_m^{(\pm)}(z, t) = \frac{k^{\frac{3}{2}}}{\sqrt{2\omega}}
~\mbox{e}^{\pm i\omega t}~\varphi_m(z)
\label{RSnonzero}
\ee
where
\be
\varphi_m (z) = z^2~
\sqrt{\frac{m}{2}}~\frac{N_1(m z_B) J_2 (m z)
- J_1(m z_B) N_2 (m z)}{\sqrt{[N_1(m z_B)]^2 +
 [J_1(m z_B)]^2}}
\label{RSKK}
\ee
These spatial functions
are normalized as follows,
\be
  \int_{z_B}^{\infty}~\frac{d z}{z^3} \varphi_m \varphi_{m'}
= \delta (m - m')
\nonumber
\ee
the factor $k^{3/2}$ in the expressions (\ref{RSzero}) and 
(\ref{RSnonzero}) is due to
the measure in (\ref{measure}).

Soft KK gravitons have low spatial momentum {\it and}
frequency (and hence mass $m$). So, we assume in the main text
that, together with relations (\ref{peta0}) and (\ref{pk}),
their counterparts
\be
     m|\eta_0| \ll 1
\label{meta0}
\ee
and
\be
   m|\eta_0| \sinh \xi_B \ll 1
\label{mk}
\ee
are satisfied.

One way
to calculate the Bogoliubov coefficients between
{\it in}- and {\it out}-vacua is to specify
the solution (of fixed frequency) at infinite future,
evolve it backwards in time and find its decomposition in
terms of the wave functions corresponding to {\it in}-vacuum.
Since  negative frequency functions in the frame
with dS$_4$ slicing
are at the same
time negative frequency functions of adS$_5$ vacuum,
we are free to use either of these two sets as defining 
{\it in}-vacuum (provided we consider the interior of 
past light cone)

Thus, the solution of interest coincides in 
the future (flat brane) region with 
one of the RS wave functions
$\varphi$. For the solution everywhere in space-time bounded 
by the brane we write
\be
  \Phi = \varphi + \Delta \Phi
\nonumber
\ee
Clearly, the first term here does not contribute
to the Bogoliubov $v$-coefficients, relating
{\it in}- and {\it out}-functions of different frequencies
(the RS wave functions have fixed frequencies in static
adS$_5$ coordinates), 
so the quantity 
of interest
is $\Delta \Phi$. Using Green's formula, we write
\be
  \Delta \Phi (\eta, \xi) = - \int~D_{adv}(\eta, \xi;
\eta', \xi_B) ~[\partial_{\xi'} \varphi (\eta', \xi')]_{\xi'=\xi_B}~
d\eta'
\nonumber
\ee
The integration here is performed along the de Sitter
brane, i.e. $ \eta' < \eta_0$. The extra term
$\Delta \Phi$ concentrates in the past light cone, so we may
use the set of the wave functions obtained with dS$_4$ slicing,
eqs.~(\ref{dSzeromode}) and
(\ref{dSfull}), as the set defining the
{\it in}-vacuum.

The advanced propagator obeys
\be
\left[\left( \eta^2 \frac{\partial^2}{\partial \eta^2}
- 2\eta \frac{\partial}{\partial \eta} + p^2 \eta^2
\right) - {\cal O}_{\xi} \right]  D_{adv}(\eta, \xi; \eta', \xi')
= \delta(\xi - \xi') \delta (\eta - \eta')
\label{prop}
\ee
where
\be
{\cal O}_\xi = \frac{\partial^2}{\partial \xi^2}
-3 \frac{\cosh \xi}{\sinh \xi} \frac{\partial}{\partial \xi}
\nonumber
\ee
The form of eq.(\ref{prop}) is chosen in such a way
that the Geen's formula holds without any additional measure
(in particular, the operator $\partial_\xi^2$  enters
eq.(\ref{prop}) without any pre-factor).

Taking into account that
 the operator ${\cal O}_\xi$ is Hermitean
(on wave functions satisfying the Neumann condition at
 $\xi=\xi_B$) with weight $\sinh^{-3} \xi $ we obtain
\be
 D_{adv}(\eta,\xi; \eta',\xi')
= \sum_{\kappa} \chi_{\kappa}(\xi) \chi_{\kappa}(\xi') 
\frac{1}{\sinh^3 \xi'} D_{\kappa}(\eta, \eta')
\nonumber
\ee
where 
\be
   D_{\kappa}(\eta, \eta')
= i \frac{1}{k^3}~\theta(\eta' - \eta) 
[\psi_{\kappa}^{(+)}(\eta) \psi_{\kappa}^{(-)}(\eta')
- \psi_{\kappa}^{(-)}(\eta) \psi_{\kappa}^{(+)}(\eta')]~
\frac{1}{\eta^{\prime 4}}
\nonumber
\ee
Here $\chi_{\kappa}$ and $\psi_{\kappa}^{(\pm)}$ are space and
time components of the graviton wave functions (including the 
zero mode)
in the frame with dS$_4$ slicing, see eqs.~(\ref{dSzeromode})
and (\ref{dSfull}).

Collecting all factors, we write
\begin{eqnarray}
\Delta \Phi (\eta, \xi)
= - \frac{i}{k^3} ~\sum_{\kappa}~\int & &
\frac{d\eta'}{\sinh^3 \xi_B \eta^{\prime 4}} 
~\theta(\eta' - \eta)~\times
\nonumber \\
& &
[\phi_{\kappa}^{(+)}(\eta, \xi) \phi_{\kappa}^{(-)}(\eta', \xi_B)
- \phi_{\kappa}^{(-)}(\eta, \xi) \phi_{\kappa}^{(+)}(\eta', \xi_B) ]
\times
\nonumber \\
& &
~[\partial_{\xi'} \varphi (\eta', \xi')]_{\xi'=\xi_B}
\label{great}
\end{eqnarray}
where $\phi_{\kappa}^{(\pm)}(\eta, \xi)$ are the total
wave functions in the frame with dS$_4$ slicing. These
are given by expressions (\ref{dSzeromode})
and (\ref{dSfull}).

The advantage of this formula is that the Bogoliubov 
coefficients follow immediately. For example, take
$\varphi = \varphi_m^{(-)}$. Then the $v$-coefficient,
that determines the creation of gravitons in this mode from
the zero-point fluctuations in the mode  $\phi_{\kappa}$,
is the coefficient in front of $\phi_{\kappa}^{(+)}(\eta, \xi)$ in
eq.(\ref{great}). Namely, we take the limit $\eta \to -\infty$
and obtain
\be
  v_{m,\kappa}=
- \frac{i}{k^3}~\int~
\frac{d\eta'}{\sinh^3 \xi_B \eta^{\prime 4}}
\phi_{\kappa}^{(-)}(\eta',\xi_B)
~[\partial_{\xi'} \varphi^{(-)}_m (\eta', \xi')]_{\xi'=\xi_B}
\nonumber
\ee
or explicitly (and omitting unnecessary primes)
\be
 v_{m,\kappa}= - \frac{i}{k^3}
~\frac{\chi_{\kappa}(\xi_B)}{\sinh^3 \xi_B}~
\int_{-\infty}^{\eta_0}~
\frac{d\eta}{\eta^{4}}
\psi_{\kappa}^{(-)}(\eta)
~[\partial_{\xi} \varphi_m^{(-)} (\eta, \xi)]_{\xi=\xi_B}
\label{vfin}
\ee
Similarly, the $u$-coefficients relating the wave functions 
with the same signs of frequencies, are obtained as
follows,
\be
u_{m,\kappa}=
- ~\frac{i}{k^3} ~\frac{\chi_{\kappa}(\xi_B)}{\sinh^3 \xi_B}~
\int_{-\infty}^{\eta_0}~
\frac{d\eta}{\eta^{4}}
\psi_{\kappa}^{(-)}(\eta)
~[\partial_{\xi} \varphi_m^{(+)} (\eta, \xi)]_{\xi=\xi_B}
\label{ufin}
\ee
Thus, calculating the Bogoliubov coefficients
amounts  to evaluate the wave functions
at the brane position, $\chi_{\kappa}(\xi_B)$,
and calculate  integrals involving Bessel functions.

\subsection{ In-out Bogoliubov coefficients: zero mode to zero mode}

We begin with the calculation of the Bogoliubov coefficients
in the case where both initial and final wave functions are
the zero modes.
The final wave function is given by eq.~(\ref{RSzero}),
and in coordinates $(\eta, \xi)$ it is
\be
   \varphi_0^{(\pm)} = -\eta_0 ~\sinh \xi_B~
\frac{k^{\frac{3}{2}}}{\sqrt{p}} \mbox{e}^{\pm ip \eta \cosh \xi}
\label{zeroo}
\ee
whereas the initial wave function is defined
in eq.~(\ref{dSzeromode}).

Collecting all factors, we obtain from the representation
(\ref{vfin})
\be
 v_{0,0} = C_1 \eta_0 ~\int_{-\infty}^{\eta_0}~
\frac{d\eta}{\eta^3} \left(\eta - \frac{i}{p} \right)
\mbox{e}^{-ip\eta (1+ \cosh \xi_B)}
\label{tmp2}
\ee
For momenta obeying (\ref{peta0}) and (\ref{pk}), the integral
in (\ref{tmp2}) is saturated at $\eta \approx \eta_0$.
We find
\be
   v_{0,0} = \frac{i C_1}{2p \eta_0}
\label{v0000}
\ee
which means that at small $H/k$, the number of zero-mode
 gravitons created from fluctuations in the zero mode
state is
\be
          n_{0,0} =
  \frac{1}{4p^2 \eta_0^2} \; , \,\;\;\;\; 
\sinh \xi_B \equiv \frac{H}{k} \ll 1
\label{13**}
\ee
whereas at large $H/k$ one has instead
\be
          n_{0,0} =
  \frac{3}{8 p^2 \eta_0^2} \sinh \xi_B
= \frac{3}{8 p^2 \eta_0^2} \frac{H}{k} \; , \,\;\;\;\; 
\sinh \xi_B \equiv \frac{H}{k} \gg 1
\label{13*}
\ee
In a similar way we get
\be
   u_{0,0} = - \frac{i C_1}{2p \eta_0}
\nonumber
\ee
These expressions coincide with the 
result of
calculation in four dimensional theory
(see Appendix 3, eqs.~(\ref{v4}) and (\ref{u4})),
up to overall enhancement factor $C_1$.
This is in accord with Ref.~\cite{Langlois:2000ns}.
It is worth noting  that the relative phase of $u_{0,0}$
and $v_{0,0}$ coincides too (this relative phase enters
the correlators of the gravitational perturbations
in the post-inflationary epoch).

We note that when calculating the two-point
correlators of the
gravitational perturbations after inflation,
 the overall factor of
$z_B^2 k^3$ will appear due to normalization in
(\ref{RSzero}). This factor may be written as
$k (H\eta_0)^2 \equiv k a^2$. The factor $k$ here
accounts for the distinction between the five-dimensional
field and effective four-dimensional field (i.e., the
distinction between the five-dimensional gravitational constant
and four-dimensional Newton's constant \cite{RS}), whereas the
factor $a^2$ exists also in four-dimensional theory
(see Appendix 3, eq.(\ref{app2})).

We conclude that the (zero mode)--(zero mode) contribution
matches the four-dimensio\-nal calculation, up to the factor $C_1$.

\subsection{Zero mode to KK modes}

Let us now calculate the Bogoliubov coefficients
from initial zero mode to final KK modes,
which describe the creation of gravitons in continuum
from zero-point fluctuations in the zero mode
of dS$_4$ slicing.
Before performing the calculation, we point out that
 the number of created KK
gravitons is given by the integral
\be
\int~dm~|v_{m,0}|^2 = \int~\frac{dm}{m} |\sqrt{m} v_{m,0}|^2
\nonumber
\ee
Hence, the quantity of interest is
\be
   \sqrt{m} v_{m,0}
\nonumber
\ee
as it is this quantity that characterizes the
number of created particles in a decimal interval
of KK masses.

We insert (\ref{dSzeromode}) and (\ref{RSKK})
into expression (\ref{vfin})
and obtain
\be
  v_{m,0} = -\frac{i C_1}{\sinh^2 \xi_B \sqrt{2p\omega}}
~I_{m,0}
\nonumber
\ee
where 
\be
I_{m,0} = \int_{|\eta_0|}^{\infty}~
\frac{d\eta}{\eta^4}~\mbox{e}^{ip\eta}
\left( \eta + \frac{i}{p} \right)
~\left(\frac{\partial}{\partial \xi} 
[\mbox{e}^{i \omega \eta \cosh \xi}
 \varphi_m(\eta \sinh \xi)]\right)_{\xi=\xi_B}
\label{Im0}
\ee
(we changed the integration variable from $\eta$ to $-\eta$).
For $p$ and $m$ satisfying (\ref{peta0}), (\ref{pk}),
(\ref{meta0}) and (\ref{mk}), we use the
expansion of $\varphi_m$ at low values of argument,
 i.e., we insert
\be
\varphi_m(\eta \sinh \xi) = 
- z_B^2 \sqrt{\frac{m}{2}}
\label{smalll}
\ee
and obtain, taking into account that 
$z_B = |\eta_0| \sinh \xi_B$, that
\be
I_{m,0} = \frac{\omega}{2p} \sqrt{\frac{m}{2}} \sinh^3 \xi_B 
\nonumber
\ee
We get finally
\be
 \sqrt{m} v_{m,0} = \frac{-i C_1}{4}
\frac{m \sqrt{\omega}}{p^{3/2}}~\sinh \xi_B
\label{vmo}
\ee
The spectrum rises towards high momenta
(recall that $v_{0,0} \propto p^{-1}$
corresponds to the ``flat spectrum''), but 
the (zero mode)--KK contribution is
suppressed, as compared to the (zero mode)--(zero mode)
result. In terms of the
number of created particles, the suppression
factor is of order $(p |\eta_0| \sinh \xi_B)^2$
or  $(m |\eta_0| \sinh \xi_B)^2$, which is small in virtue of 
eqs.~(\ref{pk}) or (\ref{mk}).
Hence, the creation of KK gravitons from the fluctuations in
the zero mode state
is a subdominant process
for soft gravitons. We note in passing
that for gravitons of higher spatial momenta, the situation
is different: we show in Appendix 4 that
the creation of KK gravitons from the zero mode fluctuations 
dominates in that case.

\subsection{KK to zero mode}

Let us now consider KK gravitons as initial states
and zero mode as the final state.
The final zero mode is again given by eq.~(\ref{zeroo}),
while initial KK modes are described in 
subsection 2.2 and Appendix 2.

We are going to calculate the integral (\ref{vfin}).
Using explicit form of the zero mode (\ref{zeroo}),
we write
\be
v_{0,\kappa} =
-\frac{\sqrt{p|\eta_0|}}{\sinh \xi_B}\cdot \chi_{\kappa}(\xi_B)
\cdot k^{-\frac{3}{2}} \sqrt{|\eta_0|}
~\int_{|\eta_0|}^{\infty}~\frac{d\eta}{\eta^3}~
\psi^{(-)}_{\kappa}(-\eta) ~\mbox{e}^{ip\eta \cosh \xi_B}
\label{TMP1}
\ee
where we again changed the integration variable
from $\eta$ to $-\eta$.

Let us begin with the integral over conformal time
\be
I_{0,\kappa} = k^{-\frac{3}{2}} \sqrt{\eta_0}
~\int_{|\eta_0|}^{\infty}~\frac{d\eta}{\eta^3}~
\psi^{(-)}_{\kappa}(-\eta) ~\mbox{e}^{ip\eta \cosh \xi_B}
\nonumber
\ee 
With the explicit form of the time functions
given in subsection 2.2,
this integral can be written as follows,
\be
I_{0,\kappa} = \frac{\sqrt{\pi}}{2}
~\int_1^{\infty}~\frac{du}{u^{3/2}}~
~\mbox{e}^{ip|\eta_0| \cosh \xi_B \cdot u} 
\mbox{e}^{-\frac{\pi \kappa}{2}} 
H_{i\kappa}^{(1)}(p|\eta_0| \cdot u)
\label{int0KK}
\ee
At small $p|\eta_0|$ and $p|\eta_0| \sinh \xi_B$
this integral is again saturated
at its lower limit, $u=1$. We make use of the representation
\be
\mbox{e}^{-\frac{\pi \kappa}{2}} 
~ H_{i\kappa}^{(1)} (p|\eta_0| \cdot u)
=  \frac{\mbox{e}^{\frac{\pi \kappa}{2}} J_{i\kappa}(p|\eta_0| \cdot u)
- \mbox{e}^{-\frac{\pi \kappa}{2}} 
J_{-i\kappa}(p|\eta_0| \cdot  u)}{\sinh (\pi\kappa)}
\label{tmp4}
\ee
and
expand in $p\eta_0$.
We obtain
\be
I_{0,\kappa} = \frac{\sqrt{\pi}}{2} \cdot
\frac{\mbox{e}^{\pi \kappa/2}}{\sinh (\pi \kappa)}~
\frac{1}{(1/2 - i\kappa) \Gamma(1+ i\kappa)}
\cdot
\left( \frac{p|\eta_0|}{2} \right)^{i\kappa}
+ (\kappa \to -\kappa)
\nonumber
\ee
At small $\kappa$ this expression behaves as
\be
 I_{0,\kappa} = \frac{2i}{\sqrt{\pi}}\cdot
\frac{1}{\kappa} 
\sin \left[ \kappa \log\left( \frac{p|\eta_0|}{2} \right)
\right]
\; , \;\;\;\; \kappa \ll 1
\label{small-kappa}
\ee
while at large $\kappa$ it behaves as
\be
I_{0,\kappa} = \frac{{\rm e}^{i\frac{\pi}{4}}}{\sqrt{2}}\cdot
    \frac{1}{\kappa^{3/2}} {\rm e}^{i\kappa \log(p|\eta_0|/2)
+ i \kappa(1-\log\kappa)} \; , \;\;\;\; \kappa \gg 1\;.
\label{large-kappa}
\ee

It remains to use the formulas for spatial functions
$\chi_{\kappa} (\xi_B)$ to obtain the Bogoliubov coefficients.
Let us begin with the case $\sinh \xi_B \equiv H/k \ll 1$.
The contributions to the number of created zero-mode gravitons,
coming from zero-point fluctuations with small and large $\kappa$,
have different form. At $\kappa \ll \sinh^{-1} \xi_B$ 
we use
eq.~(\ref{chiyB1}) and obtain
\be
 v_{0,\kappa} = 
-\frac{1}{\sqrt{2}}\left|\frac{\frac{1}{4}+i\frac{\kappa}{2}}{\frac{3}{4}
+i\frac{\kappa}{2}}\right|\cdot
\sqrt{p|\eta_0|}~\sinh \xi_B \sqrt{\kappa \tanh \kappa} 
\cdot I_{0,\kappa} \;, \;\;\; \kappa \ll \sinh^{-1} \xi_B
\label{tmp7}
\ee
At $\kappa \gg \sinh^{-1} \xi_B$, equation~(\ref{chiyB2})
with $\cos \theta_{\kappa} = 1$
is appropriate, and we find
\be
  v_{0,\kappa} = - \frac{1}{\sqrt{\pi}}
\sqrt{p|\eta_0| \sinh \xi_B} \cdot I_{0,\kappa}
\nonumber
\ee
Making use of eqs.~(\ref{small-kappa}) and (\ref{large-kappa})
we find that the integral for the number of created
gravitons is saturated at $\kappa \ll \sinh^{-1} \xi_B$ and is 
of order
\be
   \int~d\kappa~ |v_{0,\kappa}|^2 \sim p |\eta_0| \sinh^2 \xi_B
  \; , \;\;\;\; \sinh \xi_B \ll 1
\label{tmp5}
\ee
Clearly, it is suppressed as compared to the
(zero mode)--(zero mode) contribution to the number of
created gravitons, eq.~(\ref{13**}), by a factor
$(p |\eta_0|)^3 \sinh^2 \xi_B = (p_{phys}/k)^2 \cdot (p_{phys}/H)$.

Let us now turn to the case $\sinh \xi_B \equiv H/k \gg 1$.
In this case we use eq.~ (\ref{chiyB2}) and find
\be
   v_{0, \kappa} \sim \sqrt{p |\eta_0| \sinh \xi_B} \cdot
I_{0\kappa} \cos \theta_{\kappa}
\nonumber
\ee
For $\kappa \ll 1$ and $\kappa \gg 1$, one has
$\cos \theta_\kappa = 2\kappa /3 $ and $\cos \theta_\kappa =1$,
respectively. 
Making use of eqs.~(\ref{small-kappa}) and (\ref{large-kappa})
we find that the integral for the number of created
gravitons is 
of order
\be
   \int~d\kappa~ |v_{0,\kappa}|^2 \sim p |\eta_0| \sinh \xi_B
  \; , \;\;\;\; \sinh \xi_B \gg 1
\label{tmp11}
\ee
Again, it is suppressed as compared to the
(zero mode)--(zero mode) contribution, eq.~(\ref{13*}),
but now by a factor
$(p |\eta_0|)^3 = (p_{phys}/H)^3$.

We conclude that the contribution of the zero-point fluctuations of
 the KK modes into
the production of soft gravitons in the zero mode state is negligible.

\subsection{Continuum to continuum}

Finally, we calculate the Bogoliubov coefficients
in the case where 
both initial and final wave functions belong
to coninuous spectra. We are again interested
in the quantity $\sqrt{m} v_{m, \kappa}$
(see the discussion in the beginning of 
subsection 3.4).

The expression (\ref{vfin})
now reads
\be
  v_{m, \kappa}= \frac{1}{\sqrt{2\omega}}
~\frac{\chi_{\kappa}(\xi_B)}{\sinh^3 \xi_B}~
\int_{|\eta_0|}^{\infty}~\frac{d\eta}{\eta^{5/2}}~
\mbox{e}^{-\pi\kappa/2} H_{i\kappa}^{(1)}(p\eta)
~\frac{\partial}{\partial \xi} 
[\mbox{e}^{i\omega \eta \cosh \xi} 
\varphi_m(\eta \sinh \xi)]_{\xi=\xi_B}
\label{kk2kk}
\ee
We again use the expression
(\ref{smalll}) and obtain
\be
 \sqrt{m} v_{m, \kappa} 
= -\frac{i}{2}\cdot |\eta_0|^{3/2} \chi_{\kappa}(\xi_B)~
m \sqrt{\omega}~
\int_{1}^{\infty}~\frac{d u}{u^{3/2}}~
\mbox{e}^{i\omega |\eta_0| u \cosh \xi_B}
\mbox{e}^{-\pi\kappa/2} H_{i\kappa}^{(1)}(p|\eta_0|\cdot u)
\label{tmp6}
\ee
The integral here is the same as
in eq.~(\ref{int0KK}), and it is dominated by
the lower limit of the integration.
Comparing eq.~(\ref{tmp6}) with eqs.~(\ref{TMP1}) and 
(\ref{int0KK})
we
find that at small $p$ and $m$, the
creation of KK gravitons from
KK zero-point fluctuations is further suppressed
as compared to the case of subsection 3.5.
For the number of created gravitons, the
suppression factor is
of order
$(p |\eta_0| \sinh y_B)^2$ or $(m |\eta_0| \sinh y_B)^2$.
Thus, the ``KK to KK'' contribution to the  creation
of soft gravitons
is completely negligible.

\subsection{Discussion}

We have found that the main source of created gravitons is
the initial zero-point fluctuations in the {\it zero-mode} state
existing in the frame with dS$_4$ slicing.
The contributions of initial KK modes are negligible
in all cases. As shown in Appendix 4, this general property
holds not only for soft gravitons considered in this section,
but also for gravitons of somewhat higher spatial momenta.

Soft gravitons obeying (\ref{peta0}) and (\ref{pk})
(and also (\ref{meta0}) and (\ref{mk}) for KK modes)
are created predominantly in the final zero-mode state.
This is not, however, the general property of this system: 
we find in
Appendix 4 that the situation is inverse for gravitons whose
wavelengths at the end of inflation are shorter than the
adS$_5$ radius. For the moment, the novel feature of the
inflationary brane-world model --- the creation of KK gravitons,
or, in other words, the emission of five-dimensional gravitons into
the bulk --- appears of academic interest only.

\vspace{5mm}

The authors are indebted to N.~Arkani-Hamed, P.~Binetryi, N.~Deruelle,
S.~Dubovsky, D.~Langlois and P.~Tinyakov for helpful
discussions. This work has been supported in part 
by RFBR  grant 99-02-18410, CPG and SSLSS grant 00-15-96626, CRDF grant
(award RP1-2103), Swiss Science Foundation grant 7SUPJ062239. 
The work of D.G. was supported in part also under RFBR grant
01-02-06035. The work of S.S. was supported in part also under RFBR grant
01-02-06034.

\vspace{4mm}

{\Large \bf Appendix 1. Smallness of graviton momenta.}

We consider in the main text the creation of gravitons 
whose physical momentum at the end of inflation 
is smaller than the inverse adS$_5$ radius,
\be
   p_{phys} \ll k
\nonumber
\ee
Here we argue that  only
these gravitons are of interest from the point of view
of CMB anisotropy. For this purpose, we estimate the minimum
possible {\it present} momentum $p_0$ of gravitons which at the
end of inflation have 
\[
  p_{phys} = k
\nonumber
\]
We will see that in any feasible scenario,
the present wavelength $1/p_0$ is too short
to be relevant for CMB anisotropy.

In the first place, we are dealing with gravitons which are
superhorizon by the end of inflation,
\[
   p_{phys} \ll H
\nonumber
\]
Thus, we only have to consider the case
\[
  H \gg k
\nonumber
\]
There is nothing wrong with this case, but the 
cosmological evolution is somewhat different from
the conventional four-dimensional cosmology.
As shown in Ref.~\cite{Binetruy:2000hy}, 
instead of the conventional Friedmann
equation one has in this case
\be
   H^2 = \frac{16 \pi^2}{9} \cdot \kappa_5^4 \rho^2
\nonumber
\ee
where 
 $\kappa_5^2 = M_{(5)}^{-3}$ is
the five-dimensional gravitational constant. 
With $M_{(5)}^{3}= kM_{Pl}^2$
this means 
\be
  \rho_{e.i.} \sim Hk M_{Pl}^2
\label{85}
\ee
This determines the energy density just after inflation
(hence the subscript, meaning ``end of inflation'').

Let us set an upper limit on the redshift from the end of 
inflation to today.
Consider an
extreme case of very long post-inflationary preheating,
at which the equation of state corresponds to pressureless matter.
Let us assume, again as an extreme case, that this regime
terminates only at the
 nucleosynthesis epoch, $T_{NS} \sim 10$~MeV. 
Then
\be
z_{e.i.} \equiv    \frac{a_0}{a_{e.i.}} =
\frac{a_0}{a_{NS}} \frac{a_{NS}}{a_{e.i.}} \sim
\frac{T_{NS}}{T_0}~
\left(\frac{\rho_{e.i.}}{T_{NS}^4}\right)^{1/3}
\nonumber
\ee
where the subscript $0$ refers to today. 

Now, the Hubble parameter  at inflation, $H$, 
must be smaller than
the fundamental scale $M_{(5)}$, i.e.,
\be
  H < (k M_{Pl}^2)^{1/3}
\label{A32}
\ee
In fact, a stronger bound comes from gravity waves 
themselves, as their amplitude (coming from (zero mode)
-- (zero mode) Bogoliubov coefficients, with account of
$C_1 \sim \sqrt{H/k}$) is
\be
  \delta \sim \frac{H}{M_{Pl}} \sqrt{\frac{H}{k}}
\nonumber
\ee
It must be smaller than about $10^{-3}$, so instead of
(\ref{A32}) we have
\be
 H < (\delta^2 k M_{Pl}^2)^{1/3}  
\label{A34}
\ee
where $\delta \sim 10^{-3}$.

For the present wavelength of a mode whose wave 
number just after inflation was equal to $k$ we have 
\be
l_0 \equiv \frac{1}{p_0} = \frac{z_{e.i.}}{k} =
\frac{1}{k} \frac{T_{NS}}{T_0} 
\left(\frac{k H M_{Pl}^2}{T_{NS}^4}\right)^{1/3}
\label{A35}
\ee
The smaller $k$, the larger $l_0$, even if one takes 
into account (\ref{A34}). For $k^{-1} \sim 1$~mm 
one obtains from (\ref{A34})
\be
  H < 3\cdot10^{6} ~\mbox{GeV}
\nonumber
\ee
and then from (\ref{A35}) one finds
\be
  l_0 < 10^{23}~\mbox{cm} \sim 30~\mbox{kpc}
\nonumber
\ee
As claimed, this is too short scale to 
be relevant for 
the CMB anisotropy.

In all above estimates we pushed everything
to extreme, in more realistic cases the scale $l_0$
should be  much smaller.

\vspace{4mm}

{\Large \bf Appendix 2.  Graviton modes in the 
frame with de Sitter slicing}

\vspace{3mm}

{\large \bf Small $\sinh \xi_B$ and $\kappa \sinh \xi_B$}

In terms of the variable $u=\cosh^2 \xi$, eq.(\ref{dSspace})
is the hypergeometric equation
\be
u(1-u) \partial_u^2 \chi
+\left(\frac{1}{2} + \frac{1}{2} u \right) \partial_u
\chi 
- \left(\frac{9}{16} + \frac{\kappa^2}{4}\right)\chi = 0
\label{hypch}
\ee
(the subscript in $\chi_{\kappa}$ is omitted). 
Another way to
cast eq.(\ref{dSspace}) into hypergeometric form is
to introduce the variable $v = - \sinh^2 \xi$. Then 
one has
\be
v(1-v) \partial_v^2 \chi + \left(-1 + \frac{1}{2} v
  \right) \partial_v \chi -  
\left(\frac{9}{16} + \frac{\kappa^2}{4}\right)\chi = 0
\label{hypsh}
\ee
It is then convenient to choose two linear independent
solutions as follows. The first solution is
\be
  \chi_1 = F\left( \alpha, \beta; \frac{1}{2}; 
\cosh^2 \xi \right)
\nonumber
\ee
where 
\be
  \alpha = - \frac{3}{4} + i \frac{\kappa}{2} \; , \;\;\;\;
  \beta  = - \frac{3}{4} - i \frac{\kappa}{2}
\nonumber
\ee
and
the hypergeometric function $F$ should be
evaluated at a fixed side of the branch cut\footnote{Another 
choice of the side of the branch cut would correspond 
to a linear combination of the original solution and the 
second one, $\chi_2$.}
 extending from
$1$ to $\infty$ (say,  above the real axis).
This form of the solution is clear  from the representation
(\ref{hypch}).
The form of the second  solution
comes from the representation (\ref{hypsh})
\be
   \chi_2 = \sinh^4 \xi ~F\left(\frac{1}{2} - \alpha,
\frac{1}{2} - \beta; 3; -\sinh^2 \xi \right)
\nonumber
\ee
The reason to choose this particular pair of solutions
is that they behave quite differently at small $\xi$:
 the first one behaves as
\be
  \chi_1 = 
\frac{\Gamma (1/2)}{\Gamma(1/2 - \alpha) \Gamma (1/2 - \beta)}
(1 + \alpha \beta \sinh^2 \xi + O(\sinh^4 \xi ~\log (\sinh \xi)))
\label{smally1}
\ee
whereas the second one behaves as
\be
\chi_2 = \sinh^4 \xi
\label{smally2}
\ee
As a by-product, these formulae tell that $\chi_1$ and $\chi_2$
are indeed linear independent.

We now write the required solution in the form
\be
  \chi = A_1 \chi_1 + A_2 \chi_2
\label{soln1}
\ee
and impose the Neumann condition (\ref{boundaryxi}).
We first perform the calculations assuming that
\be
   \sinh \xi_B \ll 1 \;\;\; \mbox{and} 
\;\;\;\; \kappa \sinh \xi_B \ll 1
\label{Case1}
\ee
The opposite case is easy and will be considered later.
In the case (\ref{Case1}) 
we  use the small-$\xi$ expansions
(\ref{smally1}) and (\ref{smally2}) near $\xi=\xi_B$
and obtain
\be
   A_1 = - A_2\cdot \sinh^2 \xi_B \cdot
\frac{2\Gamma (1/2 - \alpha) \Gamma (1/2 - \beta)}{\alpha \beta
\Gamma(1/2)}
\label{a1}
\ee
Note that  $A_1$ is small compared to $A_2$
for small $\sinh \xi_B$.

We now wish to calculate the normalization factor.
The following normalization of continuum modes is appropriate,
\be
\int~\frac{d \xi}{\sinh^3 \xi} \chi_{\kappa}^{*}
\chi_{\kappa'} = \delta(\kappa - \kappa')
\nonumber
\ee
As is usual for continuum spectrum,
the normalization factor is
determined by large-$\xi$ asymptotics of the 
solution.
It is straightforward to find the asymptotics of the first solution,
\begin{eqnarray}
  \chi_1 = & &
\frac{\Gamma(1/2) \Gamma(-i\kappa)}{\Gamma(-3/4 - i\kappa/2)
\Gamma(5/4 - i\kappa/2)} (\cosh \xi)^{\frac{3}{2} - i\kappa}~
\mbox{e}^{i\theta}
\nonumber \\
 &+&
\frac{\Gamma(1/2) \Gamma(i\kappa)}{\Gamma(-3/4 + i\kappa/2)
\Gamma(5/4 + i\kappa/2)} (\cosh \xi)^{\frac{3}{2} + i\kappa}~
\mbox{e}^{i\theta'}
\nonumber
\end{eqnarray}
where phases $\theta$ and $\theta'$ are not important for our
purposes. The asymptotics of the second solution is
\begin{eqnarray}
  \chi_2 = & &
\frac{\Gamma(3) \Gamma(-i\kappa)}{\Gamma(5/4 - i\kappa/2)
\Gamma(7/4 - i\kappa/2)} (\sinh \xi)^{\frac{3}{2} - i\kappa}
\nonumber \\
 &+&
\frac{\Gamma(3) \Gamma(i\kappa)}{\Gamma(5/4 + i\kappa/2)
\Gamma(7/4 + i\kappa/2)} (\sinh \xi)^{\frac{3}{2} + i\kappa}
\label{aschi2}
\end{eqnarray}
At large $\xi$ one neglects $A_1 \chi_1$
term in (\ref{soln1}). Making use of (\ref{aschi2}),
one finds 
\be
  A_2 = -\frac{1}{2\sqrt{\pi}}\left| \frac{\Gamma(5/4 + i\kappa/2)
\Gamma(7/4 + i\kappa/2)}{\Gamma(3) \Gamma(i\kappa)} \right|
\nonumber
\ee
(sign here is chosen for future convenience).
This expression can be simplified by making use of
the identities,
\begin{eqnarray}
\left| \Gamma\left(\frac{5}{4} + i\frac{\kappa}{2}\right)
\Gamma\left(\frac{7}{4} + i\frac{\kappa}{2}\right)\right|^2
&=&\left| \left(\frac{1}{4} + i \frac{\kappa}{2} \right)
\Gamma\left(1 -\frac{3}{4} + i\frac{\kappa}{2}\right)
\Gamma\left(1 +\frac{3}{4} + i\frac{\kappa}{2}\right)\right|^2  
\nonumber \\
&=& \frac{2 \pi^2}{\cosh (\pi \kappa)}
\left| \frac{1}{4} + i \frac{\kappa}{2} \right|^2
\left| \frac{3}{4} + i \frac{\kappa}{2} \right|^2
\nonumber
\end{eqnarray}
\be
   |\Gamma(i\kappa)|^2 = \frac{\pi}{\kappa \sinh (\pi \kappa)}
\nonumber
\ee
We find
\be
   A_2 = -\frac{1}{2} \sqrt{\frac{\kappa}{2} \tanh \kappa}
\left| \frac{1}{4} + i \frac{\kappa}{2} \right|
\left| \frac{3}{4} + i \frac{\kappa}{2} \right|
\nonumber
\ee
For the calculation of the Bogoliubov coefficients
 we  need the value of the wave function
at the brane position, $\xi=\xi_B$.
Since 
$A_2 \chi_2 (\xi_B) \sim A_2 \sinh^4 \xi_B$ and 
$A_1 \chi_1 (\xi_B) \sim  A_2 \kappa^{-2} \sinh^2 \xi_B$,
the part $A_1 \chi_1$ is dominant on the brane.
Making use of eqs.(\ref{smally1}) and (\ref{a1})
we find
\begin{eqnarray}
  \chi_{\kappa}(\xi_B) &=& A_1 \chi_1 (\xi_B)
\nonumber \\
 &=& - A_2 \frac{2}{\left|\frac{3}{4} + i \frac{\kappa}{2}\right|^2}
\cdot \sinh^2 \xi_B
\nonumber \\
&=&
\frac{1}{\sqrt{2}}\sqrt{\kappa \tanh \kappa} 
\frac{\left|\frac{1}{4} + i \frac{\kappa}{2}\right|}{\left|\frac{3}{4} 
+ i \frac{\kappa}{2}\right|} \cdot \sinh^2 \xi_B
\label{chiyB1}
\end{eqnarray}
We recall that these formulae are valid provided the relations
(\ref{Case1}) are satisfied.
We now turn to the opposite case.

\vspace{3mm}

{\large \bf Large $\sinh \xi_B$ and/or $\kappa \sinh \xi_B$}

In the opposite case,
\be
 \sinh \xi_B \gg 1 \;\;\; \mbox{and/or} \;\;\;\; 
\kappa \sinh \xi_B \gg 1
\label{Case2}
\ee
it is convenient to introduce the wave function
\be
   \tilde{\chi} (\xi) = \chi (\xi)~\sinh^{-3/2} \xi 
\nonumber
\ee
and cast equation (\ref{dSspace}) in  the form of the 
Schr\"odinger
equation
\be
   \left( - \partial_\xi^2  + \frac{15}{4 \sinh^2 \xi}
\right) \tilde{\chi} = \kappa^2 \tilde{\chi}
\label{Schr}
\ee
In the  case of interest, 
the potential term in eq.(\ref{Schr}) may be neglected,
and the solution (normalized to delta-function)
becomes
\be
  \chi = \frac{1}{\sqrt{\pi}}
\sinh^{3/2} \xi ~\cos [\kappa(\xi  - \xi_B) + \theta_\kappa]
\nonumber
\ee
The phase here is determined from the Neumann boundary
condition on the brane, eq.(\ref{boundaryxi}). One finds
\be
   \tan \theta_\kappa = \frac{3 \cosh \xi_B}{2 \kappa \sinh \xi_B}
\nonumber
\ee
Thus, the value of the
wave function at the brane position is
\be
   \chi_\kappa (\xi_B) = \frac{1}{\sqrt{\pi}}
\sinh^{3/2} \xi_B \cos  \theta_\kappa
\label{chiyB2}
\ee
Note that for parameters obeying eq.~(\ref{Case2}), the value of the
wave function on the brane is always smaller than 
$\frac{1}{\sqrt{\pi}}\sinh^{3/2} \xi_B$.

\vspace{4mm}

{\Large \bf Appendix 3. Bogoliubov coefficients in four-dimen\-sio\-nal 
theory.}

In the four-dimensional theory one considers the metric which is 
de Sitter
at
inflationary stage
\be
   ds^2 = \frac{1}{(H\eta)^2}(d\eta^2 - d{\bf x}^2)
\; , \;\;\;\; \eta < \eta_0
\nonumber
\ee
and flat afterwards,
\be
   ds^2 = \frac{1}{(H\eta_0)^2}(d\eta^2 - d{\bf x}^2)
\; , \;\;\;\; \eta > \eta_0
\nonumber
\ee
Properly normalized solutions in the de Sitter and flat
regions are
\be
   \phi^{(\pm)} = \frac{H}{\sqrt{2p}} \mbox{e}^{\pm ip\eta}
\left( \eta \pm \frac{i}{p} \right)
\nonumber
\ee
and
\be
\varphi^{(-)} =
\frac{H}{\sqrt{2p}} \eta_0 \mbox{e}^{-ip\eta}
\label{app2}
\ee
respectively.

In the de Sitter region, one writes the solution
that becomes $\varphi^{(-)}$ at late times as a
linear combination
\be
  \Phi = u^{*} \phi^{(-)} + v \phi^{(+)}
\nonumber
\ee
and finds the Bogoliubov coefficients by matching
this solution and its derivative to $\varphi^{(-)}$
at $\eta = \eta_0$. In this way one obtains
\be
  v = \mbox{e}^{-2ip\eta_0} \frac{i}{2p\eta_0}
\nonumber
\ee
\be
  u = 1 - \frac{i}{2p\eta_0}
\nonumber
\ee
At small $p \eta_0$ (physical momentum just after inflation
small comared to $H$), these become
\be
  v = \frac{i}{2p\eta_0}
\label{v4}
\ee
\be
  u =  - \frac{i}{2p\eta_0}
\label{u4}
\ee
This is the standard ``flat spectrum''.

\vspace{4mm}

{\Large \bf Appendix 4. Creation of gravitons of higher
momenta}

At $H \gg k$, i.e. $\sinh \xi_B \gg 1$, there exists a range 
of momenta in which the gravitons are superhorizon
at the end of inflation, 
\[
       p |\eta_0| \ll 1
\nonumber
\]
and yet their physical momentum at that time exceeds the inverse
adS$_5$ radius,
\be
   p |\eta_0| \sinh \xi_B \gg 1
\label{A4*}
\ee
Here we present for completeness the results of calculations
of the Bogoliubov coefficients in this case. For 
the masses of final KK
gravitons we assume also
\be
    m |\eta_0| \ll 1 \; , \;\;\;\;
 m |\eta_0| \sinh \xi_B \gg 1
\nonumber
\ee
so that their physical frequency is small compared to
$H$ and large compared to $k$ at the end of inflation.

{\bf Zero mode to zero mode}

The integral (\ref{tmp2}) is equal to
\be
v_{0,0}={C_1\over
p\eta_0}~{\e^{i\eta_0p\cosh\xi_B}\over
p\eta_0\cosh\xi_B}
\label{v00n}
\ee
where, as before, $C_1 = \sqrt{3 \sinh \xi_B /2}$ at large
$\sinh \xi_B$. Comparing this result with eq.~(\ref{v0000})
and recalling (\ref{A4*}), we see that the spectrum of created
gravitons is damped at higher momenta
due to the extra factor 
$(p\eta_0\cosh\xi_B)^{-2}$ appearing
in $n_{0,0} \equiv |v_{0,0}|^2$.

{\bf Zero mode to KK modes}

To evaluate the integral~(\ref{Im0}), we
 use the approximation 
\bea
\frac{\partial}{\partial \xi} 
[\mbox{e}^{i \omega \eta \cosh \xi}
 \varphi_m(\eta \sinh
\xi)]\biggr|_{\xi=\xi_B}=\l\eta\sinh\xi_B\r^{5/2}
{\e^{i\omega\eta\cosh\xi_B}\over\sqrt{\pi}}
\nonumber \\
\times\l 
i\omega\cos\l m(|\eta_0|-\eta)\sinh\xi_B\r+m\coth\xi_B\sin\l m(|\eta_0|-
\eta)\sinh\xi_B\r\r
\label{derivative}
\eea
We obtain
\be
\sqrt{m}v_{m,0}=-{\sqrt{3}\over2\sqrt{2\pi}}
\coth\xi_B\e^{i|\eta_0|\omega\cosh\xi_B}{\sqrt{m\eta_0}\over 
\eta_0^2(\omega^2\coth^2\xi_B-m^2)}\sqrt{p\over\omega}
\label{vmon}
\ee
This result is to be compared with eq.~(\ref{vmo}). We see that
the spectrum no longer increases towards high momenta;
instead, it turns down at large $p$ and $m$. 
It is of interest also to compare eqs.~(\ref{vmon}) and 
(\ref{v00n}) and find that for higher momentum gravitons
\[
   \frac{|v_{0,0}|^2}{|\sqrt{m} v_{m,0}|^2}
\sim \frac{1}{p |\eta_0| \cosh \xi_B}
\nonumber
\]
This means that as far as the zero-mode initial fluctuations 
are concerned, the production of KK gravitons dominates
over the production of the zero-mode gravitons in the range of 
momenta considered in this Appendix. This is in contrast to
the case studied in the main text.

{\bf KK to zero mode}

The Bogoliubov coefficient reads
$$
v_{0,\kappa}=-{\sqrt{p|\eta_0|}\over\sinh\xi_B}\cdot\chi_\kappa(\xi_B)\cdot
I_{0,\kappa}
$$
where $\chi_{\kappa}(\xi_B)$ is given by Eq.~(\ref{chiyB2}) and 
$$
I_{0,\kappa}= \frac{\sqrt{\pi}}{2}
~\int_1^{\infty}~\frac{du}{u^{3/2}}~
~\mbox{e}^{ip|\eta_0| \cosh \xi_B \cdot u} 
\mbox{e}^{-\frac{\pi \kappa}{2}} 
H_{i\kappa}^{(1)}(p|\eta_0| \cdot u)
$$
This integral  is again saturated  at the lower
limit of integration,  so in the integrand 
we substitute the expression
for the Hankel function at small value of its argument. 
In this way we obtain
$$
I_{0,\kappa}=i{\sqrt{\pi}\over 2}{\e^{ip|\eta_0|\cosh\xi_B}\over
\sinh(\pi\kappa)}{1\over p|\eta_0|\cosh\xi_B}\l\l{p|\eta_0| 
\over2}\r^{i\kappa}{\e^{\pi\kappa\over2}\over\Gamma(1+i\kappa)} -
(\kappa\to-\kappa)\r  
$$  
The final estimate for the number of created gravitons is
\begin{equation}
\int~d\kappa~|v_{0,\kappa}|^2 \sim \l p |\eta_0| \sinh\xi_B\r^{-1}
\label{ttmp11}
\end{equation}
Comparing this estimate with eq.~(\ref{tmp11}) we find that
the spectrum is again damped at higher momenta: the ratio of
eqs.~(\ref{ttmp11}) and (\ref{tmp11}) is of order 
$(p|\eta_0|\sinh\xi_B)^{-2}$. 
Also, it follows from eqs.~(\ref{v00n}) and (\ref{ttmp11})
that 
the production of the zero mode gravitons
from KK modes is suppressed by 
the factor  $(p|\eta_0|)^3$ in comparison with the production of 
zero mode gravitons from the fluctuations in the zero mode.
This suppression is the same as for soft gravitons considered
in subsection 3.5. 

{\bf Continuum to continuum}

The Bogoliubov coefficients are given by 
eq.~(\ref{kk2kk}) where $\chi_\kappa(\xi_B)$ is determined by
\linebreak eq.~(\ref{chiyB2}), and
the approximation (\ref{derivative}) can be used. 
Thus, one has to calculate the integral
\begin{eqnarray}
v_{m, \kappa}={|\eta_0|\over
\pi}\sinh\xi_B\cos\theta_\kappa{e^{-\pi\kappa/2} \over\sqrt{\omega}}
\int_{1}^{\infty}~d 
u~H_{i\kappa}^{(1)}(p|\eta_0|\cdot
u)\e^{i\omega|\eta_0|\cosh\xi_B\cdot u}
\nonumber \\
\times\l
i\omega\cos\l m|\eta_0|(1-u)\sinh\xi_B\r+m\coth\xi_B\sin\l
m|\eta_0|(1-u)\sinh\xi_B\r\r
\nonumber
\end{eqnarray}
It is saturated again at the lower
limit of integration. We find
\begin{eqnarray}
v_{m, \kappa}&=&-{1\over2}{\cos\theta_\kappa\over\sqrt{2\omega}}
{\e^{i\omega|\eta_0|\cosh\xi_B}\over\sinh\pi\kappa}
\nonumber \\
&\times&
\Biggl\{ {\omega-m\coth\xi_B\over\omega\coth\xi_B-m}~\Biggl[ 
{\e^{\pi\kappa/2}\over\Gamma(1+i\kappa)}\l{p|\eta_0|\over2}\r^{i\kappa}\l
1-{\kappa\over|\eta_0|(\omega\cosh\xi_B-m\sinh\xi_B)}\r
\nonumber\\ &-&
(\kappa\to-\kappa)\Biggr]+(m\to-m)\Biggr\}
\label{137}
\end{eqnarray}
One can show that this expression leads to the number of created
gravitons suppressed by a factor $(p |\eta_0| \sinh \xi_B)^{-3}$
as compared to the case of soft gravitons, see eq.~(\ref{tmp6}).
Furthermore, comparing eqs.~(\ref{137}) and (\ref{vmon}), one finds
that
the production of KK modes of higher spatial momenta
from KK modes 
is suppressed by 
a factor  $(p|\eta_0|)^{3}$ in comparison with the production of the
KK modes from the zero mode. 

{\bf Discussion}

For momenta obeying (\ref{A4*}), the dominant effect is the
creation of KK gravitons from initial fluctuations in the zero mode.
In five-dimensional language, this corresponds to the emission
of the five-dimensional gravitons into the bulk.
Let us estimate the energy per unit three-volume,
emitted into the bulk in the form of
superhorizon gravitons,
\[
  \epsilon_{superhorizon} = \frac{1}{a^4 (z_B)}~
\int~\frac{dm}{m} ~\frac{d^3 p}{(2 \pi)^3} ~\omega
|\sqrt{m} v_{m,0}|^2
\nonumber
\]
where the factor $a^{-4}(z_B)$ accounts for the fact that
$p$ and $\omega$ are conformal, rather than physical, 
three-momentum and frequency. Making use of 
eq.~(\ref{vmon}), and setting the upper limit
of integration equal to $p_{max} \sim m_{max} \sim |\eta_0|^{-1}$,
we obtain an estimate
\[
 \epsilon_{superhorizon} \sim
\frac{1}{a^4(z_B) \eta_0^4} \sim H^4
\nonumber
\]
In view of eqs.~(\ref{85}) and (\ref{A34}), we find
\[
  \frac{\epsilon_{superhorizon}}{\rho_{e.i.}} < \delta^2 \ll 1
\nonumber
\]
This means that only a small fraction of energy
leaks from the brane into the bulk in the form of
gravitons
with momenta smaller than $H$. 
The emission into the bulk is a small effect, at least
for superhorizon modes.


\end{document}